\documentstyle[12pt]{article}
\begin{document}

\def\a{\alpha}
\def\b{\beta}
\def\d{{\rm d}}
\def\p{\partial}
\def\o{\over}
\def\ie{{\it i.e.}}
\def\eg{{\it e.g.}}
\def\be{\begin{equation}}
\def\ee{\end{equation}}
\def\bea{\begin{eqnarray}}
\def\eea{\end{eqnarray}}
\def\beaa{\begin{eqnarray*}}
\def\eeaa{\end{eqnarray*}}

\def\np{Nucl. Phys. }
\def\pl{Phys. Lett. }
\def\prl{Phys. Rev. Lett. }
\def\pr{Phys. Rev. }
\def\ap{Ann. Phys. }
\def\cmp{Comm. Math. Phys. }
\def\ijmp{Int. J. Mod. Phys. }
\def\mpl{Mod. Phys. Lett. }
\def\lmp{Lett. Math. Phys. }
\def\bams{Bull. AMS }
\def\am{Ann. of Math.  }
\def\jpsc{J. Phys. Soc. Jap. }
\def\topo{Topology}
\def\kjm{Kobe J. Math. }
\def\phyrep{Phys. Rep. }
\def\am{Adv. in Math. }

\setcounter{page}{0}
\begin{flushright}
hep-th/9805051\\
DFPD98/TH/21\\
US-FT-05/98\\
May, 1998
\end{flushright}

\begin{center}
{ \Large {\bf On the WDVV Equation and $M$-Theory \\}}

\vspace{.5in}

J. M. Isidro
\end{center}

\begin{center}{\it Dipartimento di Fisica ``G. Galilei", 
Via F. Marzolo 8, 35131 Padova, Italy.\\
{\tt isidro@pd.infn.it}\\}

\end{center}

\abstract{ A wide class of Seiberg--Witten models constructed by 
$M$-theory techniques 
and described by non-hyperelliptic Riemann surfaces are shown to 
possess an associative algebra of
holomorphic differentials. This is a first step towards proving 
that also these models satisfy the
Witten--Dijkgraaf--Verlinde--Verlinde equation. In this way, 
similar results known for 
simpler Seiberg--Witten models (described by hyperelliptic Riemann 
surfaces and constructed
without recourse to $M$-theory) are extended to certain 
non-hyperelliptic cases constructed in
$M$-theory.  Our analysis reveals a connection between the 
algebra of holomorphic differentials on
the Riemann surface and the configuration of $M$-theory branes  
of the corresponding Seiberg--Witten model.}

\newpage
\setcounter{page}{1}
\renewcommand{\theequation}{1.\arabic{equation}}
\setcounter{equation}{0}

\section{Introduction}

Seiberg--Witten (SW) models in 4 and more dimensions \cite{SW} 
have received renewed attention in the
context of $M$-theory \cite{M} and geometric engineering 
\cite{GE}. The elegant techniques developed
in \cite{W} allow the construction of a much larger class of SW 
models than had been known previously
\cite{LAG}. The basic elements used in \cite{W} are certain 
configurations of Dirichlet 4-branes and
solitonic 5-branes of type IIA string theory, lifted to 
11-dimensional $M$-theory. The
generalisations of SW models so obtained lie along different 
directions. First, the gauge group $G$
need no longer be simple, and thus it can now be taken to be a 
product of several simple factors,
$G=G_1\times\cdots\times G_n$. Second, a large family of SW models 
with vanishing beta function can
be generated by the inclusion of Dirichlet 6-branes. Finally, 
upon compactification of one spatial
dimension, a class of SW models can now be constructed whose 
Coulomb branch is described
by coverings of a torus (elliptic models). In the seminal work 
of \cite{W},
the gauge group was a product of $SU(N)$ factors; see also 
\cite{RANDALL}. Orthogonal and
symplectic gauge groups, and products thereof, were studied 
in \cite{KLL,KL,BRAND} by including 4-
and 6-orientifolds. The inclusion of 6-orientifolds further 
allows one to consider matter
hypermultiplets in representations other than the fundamental
\cite{KL}.

Along different lines, a generalisation of  the 
Witten--Dijkgraaf--Verlinde--Verlinde (WDVV) 
equation of topological field theory \cite{WI,DIJK,DUBROVIN} 
has been shown to hold in the
(apparently) unrelated context of SW models. There is in fact a 
deep link between topological field
theories, integrability  and Whitham hierarchies \cite{DUBROVIN, 
DONAGI, KRI, NAKATSU, MARTINEC,
DP, KP}, on the one hand, and SW theories (in various dimensions), 
on the other. This link has been
explored more recently in \cite{GORSKI, GUKOV}, also in connection 
with the matrix model of
$M$-theory \cite{SGUKOV}; some reviews are \cite{CARROLL}. Other 
related issues that have been
studied are the structure of the exact Wilsonian effective action 
beyond the prepotential ${\cal F}$,
instanton expansions, and  properties of the beta function and the 
renormalisation group equation of
these models \cite{PADOVA}, as well as weak and strong coupling 
expansions of the prepotential
${\cal F}$ \cite{DKP}. Related interest in the properties of SW 
models derived from $M$-theory has
been recently expressed in \cite{SCH}.

{}For the simple case of a SW model with an $SU(3)$ gauge group, 
the WDVV equation satisfied by the
prepotential ${\cal F}$ was first established in \cite{BM}. This 
was done from a study of the
Picard--Fuchs (PF)  equations \cite{BRANDEIS} governing the electric 
and magnetic periods ${\bf a}$
and  ${\bf a}_D$; the latter are related to the prepotential through 
the equation ${\bf a}_D= \p
{\cal F}/\p {\bf a}$. This same approach has been undertaken  
more recently in \cite{ITO}, in order
to extend it to an arbitrary simple gauge group. An alternative 
line was developed in \cite{RUSOS},
where the WDVV equation satisfied by a wide class of SW models was 
established from an analysis of
the algebra of differential forms  on the Riemann surface describing 
the corresponding SW model; see
\cite{RUSOX} for connected topics. The WDVV equation has also been 
extended in order to include the
quantum scale $\Lambda$ of the effective gauge theory 
\cite{BERTOLDI}; related issues have been
addressed in \cite{BONELLI}. It also appears to have an 
application in the theory of
Donaldson--Witten invariants of 4-manifolds \cite{YALE}.

A common feature to all the approaches mentioned in the preceding 
paragaph is that they 
rely on a technical assumption concerning the Riemann surface 
$\Sigma_g$ that governs the SW model 
in question. Namely, the surface $\Sigma_g$ must be hyperelliptic, 
\ie, it must be a 2-fold
branched covering of the Riemann sphere ${\bf CP}^1$ \cite{FARKAS, 
FULTON}. From a physical point
of view, this corresponds to the case of a simple (classical) gauge 
group $G$, possibly including
matter hypermultiplets, but always in the fundamental representation. 
Such was the case of the 
``old" SW models, as described in \cite{SW} and \cite{LAG}. 
The advent of $M$-theory and geometric
engineering has made it possible to lift these hypotheses, 
as explained above, in order to consider
products of gauge groups, or  matter hypermultiplets in 
non-fundamental representations. However,
even in those cases where the Coulomb branch of moduli space 
continues to be described by a family
of Riemann surfaces $\Sigma_g$, the latter are typically  
non-hyperelliptic, \ie, they are $n$-fold
coverings of the Riemann sphere ${\bf CP}^1$ with $n>2$ 
\cite{FARKAS, FULTON}. It thus seems natural
to ask if the prepotentials ${\cal F}$ governing these more 
general SW models constructed in
$M$-theory continue to satisfy the WDVV equation. 

It is the purpose of this paper to answer the above question 
in the affirmative,  at least for a
large family of generalised 4-dimensional SW models to be made 
precise below. The requirement of
hyperellipticity of $\Sigma_g$ can be lifted under certain 
assumptions that appear very naturally in
an $M$-theory context. In the absence of explicit expressions 
for the PF equations of these
generalised SW models, our analysis is based on a study of the 
algebra of holomorphic 1-forms on
$\Sigma_g$, along the lines of \cite{RUSOS}. Once the algebra
has been established, the ``residue formula" \cite{RUSOS} provides 
the passage to the WDVV equation;
we reserve  a proof of such a formula for an upcoming publication 
\cite{NOS}.

This paper is organised as follows. In section 2 we briefly review 
the techniques of \cite{RUSOS}
to be applied in later sections. Section 3 is devoted to an 
analysis of the simplest
non-hyperelliptic SW models: those constructed solely with 
$M$-theory 4- and 5-branes
\cite{W} and, possibly, 4-orientifolds as well \cite{KLL,BRAND}. 
In all these cases we give an
explicit construction of an associative algebra of holomorphic 
differentials. In section 4  we
present a simple ``dictionary" that allows one to read off a number 
of properties of the Riemann
surface $\Sigma_g$ from a knowledge of the $M$-theory brane 
configuration giving rise to the SW
model in question. In retrospective, this allows one to explain 
why the algebra of differentials
holds for the models of section 3. Applying the same techniques 
we examine in section 5 two new
families of SW models whose construction requires 6-branes \cite{W} 
and/or 6-orientifolds
\cite{KL}. In neither case is it possible to define an associative 
algebra of differentials on the
surface $\Sigma_g$ following the pattern of previous 
sections. Finally, in section 6 we summarise our
work and present some concluding remarks.

\renewcommand{\theequation}{2.\arabic{equation}}
\setcounter{equation}{0}

\section{Formulation of the problem}

To begin with, let us briefly review the derivation of the 
WDVV equation following the approach of
\cite{RUSOS}.

\subsection{Non-hyperelliptic Riemann surfaces}

Consider a connected, compact Riemann surface $\Sigma_g$ of 
genus $g$. As such, it will be an
$n$-fold covering of the Riemann sphere ${\bf CP}^1$, for a 
certain $n\geq 2$. We call $v$ a local
coordinate on ${\bf CP}^1$, while $t$ will denote a local 
coordinate on $\Sigma_g$. The latter can
be understood as the vanishing locus in ${\bf CP}^2$ of an 
irreducible polynomial $F(t, v)$,
\be
F(t,v)=\sum_{j=0}^n p_j(v) \,t^{n-j}=0,
\label{eq:1}
\ee 
where the $p_j(v)$ are certain polynomials in $v$. Branching 
points are the simultaneous solutions
of the algebraic equations
\be
F(t, v)=0, \qquad F_t(t, v)=0,
\label{eq:2}
\ee
where $F_t$ denotes $\p F/\p t$.

The space $\Omega^{(1,0)}(\Sigma_g)$ of holomorphic differentials 
on $\Sigma_g$ is
$g$-dimensional; let $\{\omega_j, j=1,2,\ldots, g\}$,  denote a 
basis. Any $\Sigma_g$ with
$g\leq 2$ is necessarily hyperelliptic, and thus falls into the 
special case 
of section 2.2. For $g\geq 3$, pick three independent 1-forms 
$\phi_k\in\Omega^{(1,0)}(\Sigma_g)$,
$k=1,2,3$, and consider the algebra 
\be
\omega_i \,\omega_j = C_{ij}^k \,\omega_k \,\phi_1 + D_{ij}^k 
\,\omega_k \,\phi_2 + E_{ij}^k
\,\omega_k\,\phi_3. 
\label{eq:3}
\ee
The products $\omega_i \,\omega_j$ are not linearly independent, 
as they belong to the space
$\Omega^{(2,0)}(\Sigma_g)$ of quadratic holomorphic differentials 
on $\Sigma_g$. The latter is
$(3g-3)$-dimensional. Equation (\ref{eq:3}) expresses the
decomposition $\Omega^{(2,0)}(\Sigma_g)\simeq\Omega^{(1,0)}
(\Sigma_g)\,\cdot\,
(\phi_1+\phi_2+\phi_3)$ in a particular basis. For $i$ and 
$j$ given, there are $3g$
parameters $C_{ij}^k$, $D_{ij}^k$ and $E_{ij}^k$ to adjust 
in equation (\ref{eq:3}), minus 3 zero
modes, which matches the value of ${\rm dim}\, \Omega^{(2,0)}
(\Sigma_g)$. This exact match proves
the existence and uniqueness of the algebra (\ref{eq:3}) of 
holomorphic 1-forms on $\Sigma_g$.

In what follows we will mod out in the equation above by the 
last two terms, $D_{ij}^k \,\omega_k
\,\phi_2 + E_{ij}^k \,\omega_k\,\phi_3$. This factor algebra be 
denoted symbolically by
\be
\omega_i \,\omega_j = C_{ij}^k \,\omega_k \,\phi_1 \quad {\rm mod}
\,(\omega_k\,\phi_2,\omega_k\,\phi_3).
\label{eq:4}
\ee 
Now this factor algebra need not be associative. The condition of
associativity,
\be
0=(\omega_i\,\omega_j)\,\omega_k - \omega_i\,(\omega_j\,\omega_k)= 
(C_{ij}^l\, C_{lk}^m-C_{il}^m\,
C_{jk}^l)\, \omega_m\, (\phi_1)^2\quad {\rm mod}(\omega_k\,\phi_2, 
\omega_k\,\phi_3),
\label{eq:5}
\ee
is equivalent to the statement that the matrices $C_i$ whose 
$(j,k)$ entries are $C_{ij}^k$ commute:
\be
[C_i, C_j]=0.
\label{eq:4a}
\ee
In the same vein as above one can perform a counting of the free 
parameters in
equation (\ref{eq:5}) and compare it with ${\rm dim}\,\Omega^{(3,0)}
(\Sigma_g)=5g-5$, the dimension
of the space of cubic holomorphic differentials on $\Sigma_g$. It 
turns out \cite {RUSOS} that the
number of free parameters to be adjusted is $6g-8$, which does not 
match ${\rm
dim}\,\Omega^{(3,0)}(\Sigma_g)$. So, in general, associativity 
breaks down, unless there is some
special reason for it to survive.

\subsection{Hyperelliptic Riemann surfaces}

There are special cases when one can still define an associative 
factor algebra of holomorphic
differentials. The resulting algebra will be similar, but not 
exactly equal, to that in equation
(\ref{eq:4}). One such case is that in which $\Sigma_g$ is 
hyperelliptic, \ie, when the number 
of sheets in the covering is 2.  For these surfaces we have 
\cite {FARKAS,FULTON}, after a
suitable change of variables in equation (\ref{eq:1}),
\be  
t^2=p(v)=\sum_{i=0}^{2g+2} u_i \,v^{2g+2-i}.
\label{eq:6}
\ee 
An explicit basis of $\Omega^{(1,0)}(\Sigma_g)$ is given by 
the holomorphic 1-forms
\be
\omega_j= {v^j\o t}\,\d v, \qquad j=0, 1, \ldots, g-1.
\label{eq:7}
\ee
{}From equation (\ref{eq:6})  it is obvious that $\sigma: (t, v)
\rightarrow (-t, v)$ is
an involution of $\Sigma_g$. We use the subindices $+$ and $-$ to 
denote the even and odd subspaces
of $\Omega^{(n,0)}(\Sigma_g)$, for $n=1, 2$ and $3$. Equation 
(\ref{eq:7}) implies that
$\Omega^{(1,0)}(\Sigma_g)= \Omega^{(1,0)}_{-}(\Sigma_g)$, \ie, 
all holomorphic 1-forms are odd
under $\sigma$. We now set $E_{ij}^k=0$ in equation (\ref{eq:3}), 
and {\it define} the algebra of
differentials through
\be
\omega_i\, \omega_j = C_{ij}^k \,\omega_k \,\phi_1\quad {\rm mod}
\,(\omega_k\,\phi_2),
\label{eq:8}
\ee
where $\phi_1, \phi_2\in \Omega^{(1,0)}_{-}(\Sigma_g)$. Now the 
multiplication operation
takes $\Omega^{(1,0)}_{-}(\Sigma_g)$ into $\Omega^{(2,0)}_{+}
(\Sigma_g)$, whose dimension is $2g-1$.
Further multiplication by $\Omega^{(1,0)}_{-}(\Sigma_g)$ takes us 
into
$\Omega^{(3,0)}_{-}(\Sigma_g)$, whose dimension is
$3g-2$. One can check \cite{RUSOS} that these dimensions exactly 
match the number of free
parameters required  to define an associative algebra in equation 
(\ref{eq:8}). Thus the
hyperelliptic involution $\sigma$ guarantees the existence and 
associativity of the algebra of
differentials. Finally, one can reexpress the associativity 
condition given in equation (\ref{eq:4a})
as the WDVV equation
\cite{RUSOS}:
\be
{\cal F}_i\, {\cal F}_k^{-1}\, {\cal F}_j= {\cal F}_j\, 
{\cal F}_k^{-1}\, {\cal F}_i. 
\label{eq:9}
\ee
This proves that the WDVV equation holds in the ``old" SW models 
of \cite{SW,LAG}, as they
were all described by hyperelliptic surfaces $\Sigma_g$ when $G$ 
was a classical, simple  gauge
group.

\renewcommand{\theequation}{3.\arabic{equation}}
\setcounter{equation}{0}
\section{SW models with 4- and 5-branes}

\subsection{Unitary gauge groups}

As a first example of a non-hyperelliptic SW model, let us consider 
the product gauge group $\prod
_{\a=1}^n SU(k_{\a})$, with matter hypermultiplets transforming in 
the representation
$\sum_{\a=1}^{n-1}({\bf k}_{\a},\bar {\bf k}_{\a+1})$. As shown in 
\cite{W}, the 
configuration of $M$-theory branes that produces this model is a 
chain of $n+1$ parallel 5-branes
labelled from 0 to $n$, with $k_{\a}$ 4-branes connecting the 
$\a -1$ and $\a$-th 5-branes, for $\a=
1, \ldots, n$. No semi-infinite 4-branes are assumed at either 
end of the chain of 5-branes. The
family of surfaces $\Sigma_g$ describing the Coulomb branch of 
the moduli space of this theory is
\cite{W}
\be
F(t,v)=\sum_{\a=0}^{n+1}p_{k_{\a}}(v)\, t^{n+1-\a}=0,
\label{eq:10}
\ee
where the polynomials $p_{k_{\a}}(v)$ are given by
\be
p_{k_{\a}}(v)=\sum_{j=0}^{k_{\a}} u^{(\a)}_ j\, v^{k_{\a}-j}, 
\quad \a=0,1, \ldots, n+1,
\label{eq:11}
\ee
and the genus is 
\be
g=\sum_{\a=1}^n (k_{\a}-1).
\label{eq:12}
\ee
The degrees $k_{\a}$ satisfy the condition $1<k_1\leq k_2\leq k_3
\leq \ldots \leq k_n$; this
ensures that the coefficient $b_{0, \a}$ of the 1-loop beta function 
of
$SU(k_{\a})$, $b_{0,\a}=-2k_{\a}+k_{\a+1}+k_{\a-1}$,  is negative or
zero for all $\a$. The
inexistence of semi-infinite 4-branes at either end of the chain of 
5-branes implies that
$k_0=0=k_{n+1}$. For every value of $\a$, the leading coefficient 
$u^{(\a)}_0$ of $p_{k_{\a}}(v)$ is
identified with the gauge coupling constant of the factor 
$SU(k_{\a})$, while $u^{(\a)}_1$
determines the hypermultiplet bare mass. The
$u^{(\a)}_j$ for $j=2, 3,\ldots, k_{\a}$ are a set of moduli on 
the Coulomb branch of the
$SU(k_{\a})$ factor of the gauge group.

The SW differential $\lambda_{SW}$ is given by \cite{WARNER,SPA}
\be
\lambda_{SW}=v\, {\d t\o t}.
\label{eq:14}
\ee 
Its derivatives with respect to the moduli $u^{(\a)}_j$, $j=2,3, 
\ldots, k_{\a}$ are holomorphic
on $\Sigma_g$ \cite{SW,LAG}. A straightforward computation shows 
that
\be 
\omega_j^{(\a)}=:{\p \lambda_{SW}\o\p u^{(\a)}_j}= -{1\o F_t}\, 
t^{n-\a}\, v^{k_{\a}-j}\, \d v\,+ \d
(*), 
\quad j=2,3,\ldots, k_{\a}, \quad \a= 1,2, \ldots, n,
\label{eq:15}
\ee 
and that there are $g=\sum_{\a=1}^{n}(k_{\a}-1)$  of them.

{}Following \cite{RUSOS}, we now turn to an analysis of the algebra
of the holomorphic differentials defined in equation (\ref{eq:15}).
Let $\omega_i^{(\a)}$ and $\omega_j^{(\b)}$ be given. Tentatively
we set the product $\omega_i^{(\a)}\,\omega_j^{(\b)}$ equal to
\be
\omega_i^{(\a)}\,\omega_j^{(\b)} = C_{i(\a), j(\b)}^{l(\b)}\,
\omega_l^{(\b)}\,\omega^{(\a)}\,{\rm
mod}\,{t^{2n-\a-\b}\,p'_{k_{\b}}(v)\o F_t^2}\,(\d v)^2.
\label{eq:21}
\ee 
Some comments are in order. Comparing the above with equation 
(\ref{eq:8}), we are taking $\phi_2$ in such a way that
$D_{ij}^k\,\omega_k^{(\a)}\,\phi_2= t^{2n-\a-\b}
\,p'_{k_{\b}}(v)\, (\d v)^2/(F_t)^2$. Also, the differential
$\phi_1$ of equation  (\ref{eq:8}) is now chosen to be any
$\omega^{(\a)}$ whose numerator, as a polynomial in $v$, is
coprime with $p'_{k_{\b}}(v)$; {\it any}\/ such
$\omega^{(\a)}$ will serve our purposes \cite{RUSOS}. A summation
over $l$ is implied in the above equation, but there is no
summation over $\a$ or $\b$. The
product $\omega_i^{(\a)}\,\omega_j^{(\b)}$ therefore  carries an
overall factor of $t^{2n-\a-\b}$, which is also present on the
right-hand side. We  can thus clear this common factor and
understand the remaining equation as
a polynomial in $v$. 

Let us prove that the structure constants $C_{i(\a),
j(\b)}^{l(\b)}$ are uniquely determined by equation (\ref{eq:21}).
The left-hand side is a polynomial  of degree $k_{\a}+k_{\b}-4$ in
$v$, while $p'_{k_{\b}}(v)$ has degree $k_{\b}-1$. Hence, for
equation (\ref{eq:21}) to hold,  the piece containing
$p'_{k_{\b}}(v)$ on the right-hand side must appear multiplied by a
polynomial $q_{i(\a)j(\b)}(v)$ of degree $k_{\a}-3$ in $v$.
Altogether, on the right-hand side of equation (\ref{eq:21}), the
number of coefficients to be determined is $k_{\b}-1$ (from the
structure constants $C_{i(\a), j(\b)}^{l(\b)}$), plus $k_{\a}-2$
(from the polynomial $q_{i(\a)j(\b)}(v)$), which add to a total of
$k_{\a}+k_{\b}-3$. On the other hand, identifying polynomials of
degrees
$k_{\a}+k_{\b}-4$ on both sides we have $k_{\a}+k_{\b}-3$
independent equations at our disposal. As the number of available
equations exactly matches the number of unknown coefficients, the
structure constants $C_{i(\a), j(\b)}^{l(\b)}$ defined in
(\ref{eq:21}) exist and are unique.

Equation (\ref{eq:21}) above defines a set of $k_{\a}-1$ matrices
$C_{i(\a)}$ with dimensions $(k_{\b}-1)\times (k_{\b}-1)$, the
$(j(\b), l(\b))$ entry of $C_{i(\a)}$ being equal to the structure
constant $C_{i(\a) j(\b)}^{l(\b)}$. On the other hand, the algebra
defined by equation (\ref{eq:21}) is associative, since it is a
polynomial algebra in the variable $v$. Therefore all these
$k_{\a}-1$ matrices commute among themselves, as in equation
(\ref{eq:4a}):
\be 
[C_{i(\a)}, C_{r(\a)}]=0.
\label{eq:80}
\ee

Next we observe that the right-hand side of equation (\ref{eq:21})
contains a sum over the differentials $\omega_l^{(\b)}$, for a
fixed value of $\b$. Hence it is not symmetric under the exchange of
$\a$ and $\b$. This problem does not occur when the covering is
hyperelliptic \cite{RUSOS}, since surfaces with just two sheets
correspond to a simple gauge group. We could just as well have
chosen to expand the left-hand side of equation (\ref{eq:21}) in
terms of the differentials $\omega_l^{(\a)}$, for fixed $\a$.
Repeating the above steps we can define a new set of structure
constants, $\tilde C$, through
\be
\omega_j^{(\b)}\,\omega_i^{(\a)} = \tilde C_{j{\b)},
i(\a)}^{l(\a)}\,
\omega_l^{(\a)}\,\omega^{(\b)}\,{\rm
mod}\,{t^{2n-\a-\b}\,p'_{k_{\a}}(v)\o F_t^2}\,(\d v)^2.
\label{eq:81}
\ee 
For the same reasons as above, equation (\ref{eq:81})  defines
a set of $k_{\b}-1$ matrices $\tilde C_{j(\b)}$ with dimensions
$(k_{\a}-1)\times (k_{\a}-1)$, all of which 
commute among themselves, as in equation (\ref{eq:80}):
\be  
[C_{j(\b)}, C_{s(\b)}]=0.
\label{eq:82}
\ee

We finally {\it define} the algebra of holomorphic
differentials on the non-hyperelliptic surface
$\Sigma_g$ through
\be
\omega_i^{(\a)}\,\omega_j^{(\b)} =: {1\o 2}\Big[C_{i(\a),
j(\b)}^{l(\b)}\,
\omega_l^{(\b)}\,\omega^{(\a)} +\tilde C_{j(\b),i(\a)}^{l(\a)}\,
\omega_l^{(\a)}\,\omega^{(\b)}\Big],
\label{eq:23}
\ee 
with the $C$'s and the $\tilde C$'s defined by equations
(\ref{eq:21}) and (\ref{eq:81}) above. That the algebra so defined
is associative follows from equations (\ref{eq:80}) and
(\ref{eq:82}). This definition trivially reduces to the one in
\cite{RUSOS} when the covering is hyperelliptic, and correctly
generalises the concept of an associative algebra of holomorphic
differentials to the non-hyperelliptic coverings considered in
this section. 

\subsection{Orthogonal and symplectic gauge groups}

In this subsection we will analyse some SW models whose gauge group 
is given by a product of
orthogonal and symplectic factors \cite{KLL}. In order to be
specific we will consider a gauge group
of the type $Sp(2k_1)\times SO(2k_2)\times\ldots\times 
Sp(2k_{n-1})\times SO(2k_n)$, with $n$ even.
Somewhat different (though closely related) product gauge 
groups can be described similarly; see
\cite{KLL} for details. The matter content of this theory 
will be $n-1$ half hypermultiplets 
transforming as $\sum_{\a=0}^n (2 {\bf k}_{\a}, 2 {\bf k}_{\a+1})$, 
where $2{\bf k}_{\a}$ denotes the
fundamental representation of the corresponding orthogonal or 
symplectic group $G_{\a}$. With
respect to each $G_{\a}$ there is always an even number of half 
hypermultiplets.

The brane configuration describing this model is a chain of $n+1$ 
parallel 5-branes, with a set of
$2k_{\a}$ 4-branes stretching between the 5-branes at sites $\a-1$ 
and $\a$. An orientifold 4-plane
is placed parallel to the 4-branes, in such a way that every object 
not lying on top of it must
have a mirror image. The 4-orientifold traverses the whole 
configuration at $v=0$. In particular,
between 5-branes $\a-1$ and $\a$ there will be $k_{\a}$ 4-branes 
``above" the 4-orientifold, and
another $k_{\a}$ ``below" it. No semi-infinite 4-branes are assumed
at the ends of the
configuration.

The family of surfaces $\Sigma_g$ describing the Coulomb branch of 
the moduli space 
of this theory is \cite{KLL}
\be
F(t, v)=\sum_{\a=0}^{n/2}p_{k_{2\a}}(v^2)\,
t^{n+1-2\a}+\sum_{\a=0}^{n/2}\Big[v^2\,p_{k_{2\a+1}}(v^2)+c_{2\a+1}
\Big]\, t^{n-2\a},
\label{eq:50}
\ee
where the polynomials $p_{k_{\a}}(v^2)$ are given by
\be
p_{k_{\a}}(v^2)=\sum_{j=0}^{k_{\a}}u_{2j}^{(\a)}\,v^{2k_{\a}-2j},
\qquad \a=0,1,\ldots,n+1,
\label{eq:51}
\ee
and the $c_{2\a+1}$ are certain numerical constants irrelevant for 
our purposes. The inexistence of
semi-infinite 4-branes at either end of the brane configuration 
implies that $k_0=0=k_{n+1}$, while
the $k_{\a}$ when $\a=1,\ldots, n$ satisfy a constraint imposed by 
the requirement of asymptotic
freedom. Namely, let  $q_{\a}=(-1)^{\a+1}$ denote the charge of the 
4-orientifold ``to the left" of
the 5-brane at site $\a$. Then the coefficient $b_{0,\a}$ of the 
1-loop beta function of the group
factor $G_{\a}$ is proportional to $a_{\a}-a_{\a-1}$, where $a_{\a}=
2k_{\a+1}-2k_{\a}-2q_{\a}$.
Asymptotic freedom therefore requires that $a_0\geq a_1\geq \dots 
\geq a_{n-1}\geq a_n$. Furthermore,
for every value of $\a$, the leading coefficient $u_0^{(\a)}$ of the 
polynomial $p_{k_{\a}}(v^2)$
is interpreted as the gauge coupling constant of the group factor 
$G_{\a}$, while the $u_{2j}^{(\a)}$ for $j=1,2,\ldots, k_{\a}$ are a 
set of moduli on the Coulomb
branch of moduli space. Contrary to the case of unitary gauge 
groups, all hypermultiplet bare masses
are zero \cite{KLL}. 

The genus $g$ of the family of surfaces $\Sigma_g$ defined in 
equations (\ref{eq:50}) and
(\ref{eq:51}) can be easily computed with the aid of the 
Riemann--Hurwitz formula
\cite{FARKAS, FULTON}. One finds
\be
g=\sum_{\a=1}^n(2k_{\a}-1).
\label{eq:52}
\ee
This value of $g$ is greater than $\sum_{\a=1}^n k_{\a}$, which 
is the number of independent
moduli (\ie, the dimension of the Coulomb branch or, equivalently, 
the rank of the product
gauge group).  As explained in \cite{KLL}, one must restrict to a 
subvariety of the full
Jacobian \cite{FARKAS, FULTON} of $\Sigma_g$ in order to obtain 
physically meaningful values for the
electric and magnetic periods ${\bf a}$ and ${\bf a}_D$  entering 
the BPS mass formula. This
subvariety is the so-called Prym variety \cite{MARTINEC} of 
$\Sigma_g$, whose dimension is
$2\sum_{\a=1}^n k_{\a}$, \ie, twice that of the Coulomb branch of 
moduli space.

In fact, the surface $\Sigma_g$ defined by equations (\ref{eq:50}) 
and (\ref{eq:51}) possesses 
an involution $\sigma\colon (v,t)\rightarrow (-v,t)$. The SW 
differential given in equation
(\ref{eq:14}) is odd under this involution, \ie, 
$\sigma(\lambda_{SW})=-\lambda_{SW}$. Hence the
holomorphic differentials generated by modular differentiation of 
$\lambda_{SW}$ will also be odd
under $\sigma$. One finds using equations (\ref{eq:50}) and 
(\ref{eq:51}) 
\bea 
\omega_j^{(2\a+1)}&=:&{\p \lambda_{SW}\o \p u_j^{(2\a+1)}}=
-{1\o F_t}\,
t^{n-1-2\a}\,v^{2k_{2\a+1}+2-2j}\,\d v + \d (*),\quad  j=1,2,
\ldots, k_{2\a+1}\cr
\omega_j^{(2\a)}&=:&{\p \lambda_{SW}\o \p u_j^{(2\a)}}=-{1\o F_t}
\, t^{n-2\a}\,v^{2k_{2\a}-2j}\,\d v
+ \d (*), \qquad j=1,2,\ldots, k_{2\a}. 
\label{eq:53}
\eea
The above differentials span a basis of the subspace of $\sigma$-odd 
holomorphic 1-forms. The
dimension of the latter is $\sum_{\a=1}^n k_{\a}$.  These are the 
differentials that are to be
integrated in order to construct the Prym variety.

Again following \cite{RUSOS}, we now turn to an analysis of the 
algebra of the holomorphic differentials in equation (\ref{eq:53}).
We have to define the products $\omega_i^{(2\a)}\, 
\omega_j^{(2\b)}$, $\omega_i^{(2\a)}\, \omega_j^{(2\b+1)}$ and
$\omega_i^{(2\a+1)}\, \omega_j^{(2\b+1)}$. These three cases must
be studied separately, although the conclusions turn out to be
the same, so we will just present the details pertaining to the
case of  $\omega_i^{(2\a)}\, \omega_j^{(2\b)}$. We
tentatively define it through
\be
\omega_i^{(2\a)}\,\omega_j^{(2\b)} = C_{i(2\a), j(2\b)}^{l(2\b)}\,
\omega_l^{(2\b)}\,\omega^{(2\a)}\,{\rm
mod}\,{t^{2n-2\a-2\b}\,p'_{k_{2\b}}(v^2)\o F_t^2}\,(\d v)^2,
\label{eq:90}
\ee 
where, as in the previous subsection, a summation is implied over
$l$, but not over $\a$ nor $\b$. Also, $\omega^{(2\a)}$ 
can be taken to be any
differential whose numerator, as a polynomial in $v$,  is coprime
with
$p'_{k_{2\b}}(v^2)$.  The product
$\omega_i^{(2\a)}\,\omega_j^{(2\b)}$  carries an overall 
factor of $t^{2n-2\a-2\b}\,(\d
v)^2/(F_t)^2$, which is also present on the right-hand side. We  
can clear this common factor
and understand the remaining equation  as a polynomial in $v$. 

We first observe that the left-hand side is a polynomial of degree 
$2k_{2\a}+2k_{2\b}-4$ in $v$, while
$p'_{k_{2\b}}(v^2)$ has degree $2k_{2\b}-1$.  Hence, for equation 
(\ref{eq:90}) to hold, 
the piece containing $p'_{k_{2\b}}(v^2)$ on the right-hand side
must  appear multiplied by a
polynomial $q_{i(2\a)j(2\b)}(v)$ of degree $2k_{2\a}-3$ in $v$.
Altogether, on the right-hand side of equation
(\ref{eq:90}), the number of coefficients to be determined add to a 
total of $2k_{2\a}+2k_{2\b}-3$. There is
a contribution of $2k_{2\a}-2$ to this quantity from the polynomial 
$q_{i(2\a)j(2\b)}(v)$, while the
structure constants $C_{i(2\a), j(2\b)}^{l(2\b)}$ contribute 
$2k_{2\b}-1$. This latter number
comes from the fact that, although there are only $k_{2\b}$ 
$\sigma$-odd differentials
$\omega_l^{(2\b)}$, one must also impose the condition that all 
$\sigma$-even terms vanish. On the
other hand, identifying polynomials of degrees
$2k_{2\a}+2k_{2\b}-4$ on  both sides we have $2k_{2\a}+2k_{2\b}-3$
independent equations at our disposal. As  the number of available 
equations exactly matches the
number of unknown coefficients, the structure constants $C_{i(2\a), 
j(2\b)}^{l(2\b)}$  exist and are unique. They define a set of
$2k_{2\a}-1$ commuting matrices $C_{i(2\a)}$ of dimensions
$(2k_{2\b}-1)\times (2k_{2\b}-1)$, as corresponds to an associative
algebra.

Next we define a new set of  structure constants $\tilde
C_{j(2\b),  i(2\a)}^{l(2\a)}$ by exchanging the indices $\a$ and
$\b$ above:
\be
\omega_j^{(2\b)}\, \omega_i^{(2\a)}\, = \tilde C_{j(2\b), i(2\a)}
^{l(2\a)}\,
\omega_l^{(2\a)}\,\omega^{(2\b)}\,{\rm
mod}\,{t^{2n-2\a-2\b}\,p'_{k_{2\a}}(v^2)\o F_t^2}\,(\d v)^2.
\label{eq:91}
\ee 
Again this defines a set of $2k_{2\b}-1$ commuting matrices $\tilde
C_{j(2\b)}$ of dimensions $(2k_{2\a}-1)\times (2k_{2\a}-1)$.
Finally,  the complete product
$\omega_i^{(2\a)}\, \omega_j^{(2\b)}$ is defined as the half-sum
of the right-hand sides of equations (\ref{eq:90}) and
(\ref{eq:91}). As in the previous subsection, the algebra so
defined is associative.

\renewcommand{\theequation}{4.\arabic{equation}}
\setcounter{equation}{0}

\section{Brane configuration and structure of $\Sigma_g$}

We have found in section 3 that an associative factor algebra of 
holomorphic differentials can be
defined on the non-hyperelliptic Riemann surfaces describing certain 
families of generalised SW
models. The algebra satisfied conforms to the pattern of equation 
(\ref{eq:8}), which was seen in
section 2 to be the algebra of hyperelliptic surfaces. There might 
seem to be an inconsistency
between the conclusions of sections 2 and 3. We devote this section 
to a resolution of this apparent
puzzle. As it turns out, there is an intimate link between the brane 
configuration that gives rise
to the SW model in question, the structure of its corresponding 
Riemann surface $\Sigma_g$, and the
possibility of defining an associative factor algebra of holomorphic 
differentials following equation
(\ref{eq:8}). For the sake of simplicity, we will concentrate for 
the rest of this section on the
case of unitary gauge groups dealt with in section 3.1. This 
dispenses with the need to project onto
a certain subspace of differentials or, equivalently, onto a certain 
Prym subvariety. However, it
will become clear that our conclusions can be easily generalised to 
all the models dealt with in the
previous section.

If $n>1$, the surface $\Sigma_g$ as defined by equations 
(\ref{eq:10}) and (\ref{eq:11}) is
non-hyperelliptic. According to \cite{RUSOS}, the algebra of 
differentials (\ref{eq:4}) always
exists, but it is not guaranteed to be associative.  However, 
the structure of
$\Sigma_g$ is such that it allows one to establish an associative 
factor algebra of
holomorphic 1-forms, in a way that closely resembles the 
hyperelliptic case of equation
(\ref{eq:8}). In fact we have already exhibited the algebra; it 
remains to explain why it can be
established. We will do so using two alternative, though 
substantially equivalent arguments. The
first one provides a dictionary that allows one to read off a number 
of properties of the
Riemann surface from the underlying brane configuration. The second 
argument, more concise, relies
on a counting of moduli.

Let us review some properties of $\Sigma_g$ from the construction 
of 
this model in \cite{W}. There
are as many sheets in the covering as there are 5-branes, so 
$\Sigma_g$ is an $(n+1)$-fold covering
of the base ${\bf CP}^1$. Every sheet of $\Sigma_g$ is a copy of 
the complex $v$-plane ${\bf C}$,
later compactified to ${\bf CP}^1$. Assume factorising $F(t,v)$ as 
$\prod_{\a=0}^{n} (t-t_{\a}(v))$.
Then $t_{\a}(v)$ is a local coordinate on the $\a$-th sheet. There 
is a branching between adjacent
sheets at sites $\a-1$ and $\a$ whenever the coordinate $v$ on the
 base ${\bf CP}^1$ is such that
$t_{\a-1}(v)=t_{\a}(v)$ for that particular value of $v$. This 
indicates the presence of a 4-brane
with coordinate $v$; there are $k_{\a}$ such values of $v$, all 
different, each corresponding to one
of the $k_{\a}$ 4-branes that stretch between sheets $\a-1$ 
and $\a$. 

No single 4-brane can connect non-adjacent sheets, \ie, sheets 
$\a-1$ and $\a-1+s$ for $s>1$.
However, a branching between sheets $\a-1$ and $\a-1+s$ for $s>1$
 can occur if $s$ 4-branes are
positioned as follows. For a fixed $v_0$ on the base ${\bf CP}^1$, 
it must hold that
$t_{\a-1}(v_0)=t_{\a}(v_0)=\ldots=t_{\a-1+s}(v_0)$. In this case, 
for every $r=0,1,\ldots, s-1$, one
4-brane out of the $k_{\a+r}$ between sheets $\a-1+r$ and $\a+r$ 
has a projection $v_0$ on the base
${\bf CP}^1$. All these $s$ 4-branes lie ``one after another", 
thus producing a branching between
$s+1$ sheets of $\Sigma_g$, with a branching index $B=s+1$.

It is clear that if non-adjacent sheets $\a-1$ and $\a-1+s$ 
for some  $s>1$ are branched together,
then it must be in the manner just described. In particular, 
{\it all}\/ intermediate sheets $\a$,
$\a+1,\ldots,\a-2+s$ will be involved in the branching; none of 
them are bypassed. It also holds
that  $v=\infty$ is not a branching point; this follows from the 
compactification of each sheet of
$\Sigma_g$  \cite{W}. We will also assume that there is no branching
 at $v=0$. 

{}For later purposes it will be instructive to compute the genus 
$g$. This we do with the aid of the
Riemann--Hurwitz formula \cite{FARKAS, FULTON}. For an $(n+1)$-fold 
covering of ${\bf CP}^1$, it
holds that
\be
\sum_{p\in \Sigma_g}(B(p)-1) = 2g + 2n,
\label{eq:12a}
\ee
where $B(p)$ denotes the branching index at point $p\in 
\Sigma_g$. The summand $(B(p)-1)$ vanishes
except at a finite number of points (branching points) \cite{FARKAS,
 FULTON}. Let us first consider 
a situation in which $B(p)=2$ at all branching points. This 
corresponds to a brane configuration in
which {\it no}\/ 4-branes lie one after another in the manner 
described above.  Hence in this
case the sum  $\sum_{p\in \Sigma_g}(B(p)-1)$ equals the total 
number of branching points. On the 
$\a$-th sheet there  are $k_{\a}$ 4-branes ``coming in" and 
$k_{\a+1}$ 4-branes ``going out", so the
total number of branching points on $\Sigma_g$ is $\sum_{\a=0}^{n}
(k_{\a}+ k_{\a+1})=
2\sum_{\a=1}^{n}k_{\a}$. From here we conclude $g=\sum_{\a=1}^{n}
(k_{\a}-1)$ as in \cite{W}. This
value of the genus stays the same if the requirement that the 
branching index be $B=2$ is lifted.
There is then a decrease in the number of branching points, but it 
is  compensated by an equal
increase in the branching index $B$. 

Now let $\Sigma_g$ be  non-singular, \ie, assume that the derivatives
 $F_t$ and $F_v$ never vanish
simultaneously on $F=0$ \cite{FULTON}, and consider the 1-forms 
given  by
\be
\phi_j^{(\a)} =: {v^{k_{\a}-j}\o \p^{n-\a+1} F/\p t^{n-\a+1}}\, 
\d v, \quad j=2,3,\ldots, k_{\a},
\quad \a= 1,2, \ldots, n.
\label{eq:16}
\ee
For $\a=n$, the holomorphicity of $\phi_j^{(n)}$ on $\Sigma_g$ when 
$j=2,3,\ldots, k_n$ follows
simply from the fact that $\phi_j^{(n)}=-\omega_j^{(n)}$, as per 
equations (\ref{eq:15}) and
(\ref{eq:16}). However, let us provide an alternative argument that 
will be useful in what follows.

The simultaneous equations $F_t=0$ and $F=0$ hold  at the branching 
points of $\Sigma_g$.   Now,
from $\d F= F_v\,\d v + F_t\, \d t =0$ and the assumption of 
non-singularity, whenever $F_t=0$ on
$F=0$ we can write $\d v / F_t = -\d t/ F_v$, with $F_v
\neq 0$. This alternative expression for
$\phi_j^{(n)}$ proves that it has no poles at finite points
 $v\neq 0$. If it has any poles at all,
then they will be at $v=0$ or $v=\infty$. In fact one can prove 
that the divisor
$[\phi_j^{(n)}]$ is given by 
\be 
[\phi_j^{(n)}]=(k_n-j)\,(0_{n-1}+0_{n}) + \Big(\sum_{\a=1}^{n-1} 
k_{\a} +
j-(n+1)\Big)\,(\infty_{n-1}+\infty_{n}),
\label{eq:17}
\ee 
where $0_{\a}$ (respectively, $\infty_{\a}$) denotes the point on 
the $\a$-th sheet of the
covering $F=0$ lying above $v=0$ (respectively, $v=\infty$) on the 
base ${\bf CP}^1$. In our
conventions, zeroes (repectively, poles) carry positive 
(respectively, negative) coefficients in the
divisor.  A proof of equation (\ref{eq:17}) is given in the 
appendix. 

Now for any meromorphic 1-form $\varphi$ on $\Sigma_g$ it holds 
that \cite{FARKAS, FULTON} 
\be
\sum_{p\in\Sigma_g}{\rm ord}_p\, (\varphi)=2g-2,
\label{eq:13}
\ee
\ie,  the zeroes minus the poles of $\varphi$  must equal 
$2g-2$. The divisor
$[\phi_j^{(n)}]$ satifies this requirement since, by equation 
(\ref{eq:12}),
\be 
2(k_n-j)+2\Big(\sum_{\a=1}^{n-1}k_{\a}+j-(n+1)\Big)=2g-2.
\label{eq:18}
\ee 
Hence $\phi_j^{(n)}$ is holomorphic on the surface $F(t,v)=0$ 
precisely when $j=2,3,\ldots, k_n$.
However, when $\a<n$, the $\phi_j^{(\a)}$ are not {\it a priori}
\/ assured to be holomorphic on
$\Sigma_g$.

Let us observe that, if the surface  $F(t,v)=0$ corresponds to the 
configuration of branes described
above, then the operation of taking the derivative $\p/ \p t$ 
corresponds to the removal of the
5-brane at site $\a=n+1$ (the one ``farthest to the right" in the 
conventions of \cite {W}). So the
surface $\p F/ \p t=0$ describes a configuration of $n$ 5-branes, 
with $k_n$ semi-infinite 4-branes
to the right of the $n$-th 5-brane. Applying the Riemann--Hurwitz 
formula of equation
(\ref{eq:12a}), its genus turns out to be $g_{n-1}=\sum_{\a=1}^{n-1} 
(k_{\a}-1)$. Similarly, the
second derivative $\p^2 F/ \p t^2=0$ describes a configuration of 
$n-1$ 5-branes, with $k_{n-1}$
semi-infinite 4-branes to the right of the $(n-1)$-th 5-brane, 
and genus 
$g_{n-2}=\sum_{\a=1}^{n-2} (k_{\a}-1)$. In general, the surface 
$\p^l F/\p t^l=0$ corresponds to  a
configuration of $(n-l+1)$ 5-branes, with $k_{n-l+1}$ semi-infinite 
4-branes to the right of the
5-brane  at site $(n-l+1)$, and genus $g_{n-l}=\sum_{\a=1}^{n-l} 
(k_{\a}-1)$. After $l$ derivatives
have been taken, the gauge group is $\prod _{\a=1}^{n-l} 
SU(k_{\a})$. For $l=n-1$ we are left with a
configuration of just two 5-branes, with $k_1$ 4-branes stretched 
across them, and $k_2$
semi-infinite 4-branes to the right: this is an $SU(k_1)$ gauge 
theory with $N_f=k_2$ fundamental
flavours. As such it is already hyperelliptic.

Consider now $\a=n-1$. The above arguments establish that 
$\phi_j^{(n-1)}$, for $j=2,3,\ldots,
k_{n-1}$,  is holomorphic  on the surface $\p F/\p t =0$, 
provided the latter is non-singular.
In general, assume that the surfaces $\p^{n-\a} F/ \p t^{n-\a}=0$ 
are non-singular for all values of
$\a=n-1, n-2,\ldots, 1$. Then, for every fixed value of $\a=n-1, 
n-2,\ldots, 1$, the 1-forms
$\phi_j^{(\a)}$ for $j=2,3,\ldots, k_{\a}$ are holomorphic on the  
surface $\p^{n-\a} F/ \p
t^{n-\a}=0$. An expression for the divisor $[\phi_j^{(\a)}]$ on the 
surface  $\p^{n-\a} F/
\p t^{n-\a}=0$ can be easily given, provided that $v=0$ is not a 
branching point on the surface 
$\p^{n-\a} F/ \p t^{n-\a}=0$. As proved in the appendix, it is given 
by
\be 
[\phi_j^{(\a)}]=(k_{\a}-j)\,(0_{\a-1}+0_{\a}) + \Big(\sum_{l=1}^
{\a-1} k_l +
j-(\a+1)\Big)\,(\infty_{\a-1}+\infty_{\a}).
\label{eq:19}
\ee 
It also satisfies the requirement of equation (\ref{eq:13}), for a 
value of the genus
$g_{\a}=\sum_{l=1}^{\a}(k_l-1)$.

If instead of considering the 1-form $\phi_j^{(\a)}$ on the surface 
$\p^{n-\a} F/ \p t^{n-\a}=0$
we consider it on $F=0$, then its divisor is given by
\be  
[\phi_j^{(\a)}]=(k_{\a}-j)\,(0_{\a-1}+0_{\a}) + \Big(\sum_{l\neq\a}
^{n} k_l +
j-(n+1)\Big)\,(\infty_{\a-1}+\infty_{\a}).
\label{eq:20}
\ee 
Of course, the actual value of the genus that will now satisfy 
the requirement of equation
(\ref{eq:13}) is $g=g_{n}=\sum_{l=1}^{n}(k_l-1)$. We observe in 
the above equation that the
coefficients of the divisor $[\phi_j^{(\a)}]$ are positive precisely 
when $j=2,3,\ldots,k_{\a}$ and
$\a=1,2,\ldots,n$; this ensures holomorphicity of the 1-forms $\phi_j
^{(\a)}$ on the surface
$F=0$. 

This completes the proof that the 1-forms given in equation 
(\ref{eq:16}) constitute a basis of
$\Omega^{(1,0)}(\Sigma_g)$, under the assumption of simultaneous 
non-singularity of the  
$\p^{n-\a} F/ \p t^{n-\a}=0$ for all $\a=n-1, n-2, \ldots, 1$. 
Following \cite{RUSOS}, it is now
immediate to establish an associative factor algebra for the 
holomorphic differentials defined in
equation (\ref{eq:16}). Let $\phi_i^{(\a)}$ and $\phi_j^{(\b)}$ 
be given. For any fixed values of
$\a$ and $\b$, we first define a set of structure constants $C$
through 
\be
\phi_i^{(\a)}\,\phi_j^{(\b)} = C_{i(\a), j(\b)}^{l(\b)}\, \phi_l
^{(\b)}\,\phi^{(\a)}\quad {\rm
mod}\,{p'_{k_{\b}}(v)\,(\d v)^2 \o (\p^{n-\a+1} F/\p
t^{n-\a+1})(\p^{n-\b+1} F/\p t^{n-\b+1})},
\label{eq:21a}
\ee 
where, as usual, a summation is implied over $l$, but not over $\a$
nor $\b$,  and we require that the numerator of 
$\phi^{(\a)}$ be coprime with $p'_{k_{\b}}(v)$. It suffices to
repeat the argument provided at the end of subsection 3.1 in order 
to prove that the above equation
uniquely defines a set of structure constants $C$. The whole
argument goes through, without the need to cancel any $t$-dependent
factors from  the differentials as done in that
subsection. The structure constants $C$ in
the above equation are actually coincident with those of
equation (\ref{eq:21}). Next one defines a new set of structure
constants
$\tilde C$, by simply exchanging $\a$ and $\b$ in equation
(\ref{eq:21a}). Finally, the complete product
$\phi_i^{(\a)}\,\phi_j^{(\b)}$ is defined as the half-sum
of the piece with the $C$'s and the piece with the
$\tilde C$'s, as in equation (\ref{eq:23}). Associativity is a
simple consequence of the fact that  the algebra itself has been
reduced to a polynomial algebra. 

{}For future reference in section 5 we make the following 
observation. In the basis of equation
(\ref{eq:15}), for any fixed value of $\a$, we span a subspace of 
differentials  by letting $j$ run
over the range $j=2,3,\ldots,k_{\a}$. We can understand this 
subspace as contributing by an amount 
$k_{\a}-1$ to the overall genus $g$ given in equation 
(\ref{eq:12}). As $\a$ runs over the range
$1,2,\ldots, n$,  we can interpret multiplication by the prefactor 
$t^{n-\a}$ for $\a=1,2,\ldots, n$
as taking us from one pair of sheets $(\a-1,\a)$ to the 
next. There is a well-defined explicit
dependence of the basis $\omega_j^{(\a)}$ on the variable $t$, 
namely, a simple monomial $t^{n-\a}$.
In passing from the  $\omega_j^{(\a)}$ to the $\phi_j^{(\a)}$ as 
a basis, we are cancelling this
explicit dependence in the numerator, at the cost of increasing 
the order of $t$-derivatives in the
denominators of the differentials. It can be done {\it  without 
losing holomorphicity}.\/ This
property of $\Sigma_g$ follows from the underlying brane 
configuration. As a consequence,
establishing the algebra of differentials in the non-hyperelliptic 
models of section 3 has been
reduced, basically, to that of $n(n-1)/2$ ``equivalent
hyperelliptic  problems". Every pair 
of adjacent sheets defines a
``hyperelliptic building block"; the structure of $\Sigma_g$ can be 
understood, roughly speaking, as
a superposition of $n$ such blocks.

This brings us to a counting of moduli, in order to  clarify the 
structure of $\Sigma_g$ as a
``superposition of hyperelliptics". The moduli of the models just 
examined are the order parameters
$u_j^{(\a)}$ on the Coulomb branch. The latter are associated with 
the gauge group $G$. Let us start
from the $SU(k_1)$ gauge theory described by a hyperelliptic surface
 with genus $g=k_1-1$
and $k_1-1$ independent moduli. The addition of the group factor 
$SU(k_2)$ to the gauge group
corresponds to adding one more sheet to the covering, with an 
additional $k_2-1$ new moduli, and a
contribution of $k_2-1$ to the genus. In general, with respect to 
the hyperelliptic case, when $G$
was a simple factor, the only new moduli that appear in these 
non-hyperelliptic models are those
associated with a product gauge group $G=G_1\times\cdots\times
 G_n$. This is an equivalent statement
of the fact, already observed, that the counting of undetermined 
parameters in the algebra of differentials, versus
that of available equations, closely resembles the hyperelliptic
case of \cite{RUSOS}.

It remains to explain, in $M$-theory terms, why any hyperelliptic 
building block admits an
associative algebra of holomorphic differentials. We saw in section 
2.2 that this can be traced
back to the existence of the hyperelliptic involution $\sigma$. The 
latter has a very natural
interpretation in $M$-theory. Namely, let us recall from \cite{W} 
that {\it the 4-brane and the
5-brane lift to one and the same basic object in $M$-theory}.\/ The 
type IIA 5-brane on ${\bf
R}^{10}$ is simply an $M$-theory 5-brane on ${\bf R}^{10} \times 
{\bf S}^1$ whose worldvolume,
roughly, is located at a point in ${\bf S}^1$ and spans a 6-manifold 
in ${\bf R}^{10}$.
A type IIA 4-brane is an $M$-theory 5-brane that is wrapped over the 
${\bf S}^1$. The type IIA
configuration of parallel 5-branes joined by 4-branes can be 
reinterpreted in $M$-theory as a
single 5-brane with a more complicated world history. It sweeps out 
arbitrary values of the first
four coordinates $x^0, x^1, x^2, x^3$ of 11-dimensional 
space--time. It is located at
$x^7=x^8=x^9=0$. In the remaining four coordinates  $x^4, x^5, x^6$  
and $x^{10}$, which parametrise
a 4-manifold ${\bf Q}\simeq {\bf R}^3\times {\bf S}^1$, the 5-brane 
worldvolume spans a
two-dimensional surface $\Sigma_g$. Then our coordinates $v$ and $t$ 
are defined \cite{W} as $v=x^4
+ {\rm i}\,x^5$ and $t=\exp [-(x^6+ {\rm i}\, x^{10})/R]$, where $R$ 
is the radius of ${\bf S}^1$.
So the hyperelliptic involution $\sigma\colon (t,v)\rightarrow 
(-t, v)$ is nothing but the statement
that, in its propagation, the $M$-theory 5-brane crosses  $x^{10}$ 
and its diametrically opposed
point $x^{10}+\pi\,R$.

In section 3 we found it convenient to use the basis given by the 
$\omega_j^{(\a)}$. This was
natural, as it was the basis obtained by straight modular 
differentiation of the SW differential
$\lambda_{SW}$. The construction given in this section by means of 
the auxiliary surfaces
$\p^{n-\a} F/\p t^{n-\a}=0$ for $\a=1,2,\ldots, n-1$ has a simple 
$M$-theory origin that highlights
the  similarities between this non-hyperelliptic case and the 
hyperelliptic surfaces dealt with in
\cite{RUSOS}. In the following section we will exhibit some new SW 
models where these similarities
cease to exist. We will again resort to their $M$-theory construction
 in order to reveal the effects
caused by the loss of these similarities.

\renewcommand{\theequation}{5.\arabic{equation}}
\setcounter{equation}{0}

\section{SW models with 6-orientifolds and 6-branes}

\subsection{Models with 6-orientifolds}

Let us now study the effect of introducing one 6-orientifold into 
the brane
configuration. Following \cite{KL}, there are basically two different
 choices to place it.
In the first one, the 6-orientifold is located between the 4-branes 
and the 5-branes, in such a way
that the resulting gauge group is of the type $\prod_{\a}SU(k_{\a})
\times SO(k_{\a})$ or
$\prod_{\a}SU(k_{\a})\times Sp(k_{\a})$, with a certain 
hypermultiplet content that typically
transforms as a sum of bifundamental (and/or vector) 
representations. We will not be interested in
these configurations. For brevity, we will be interested in placing 
the 6-orientifold on top of one
5-brane. This will bring us to interesting conclusions without 
substantial loss of generality.

Specifically, it is known \cite{KL} that a SW model with an $SU(N)$ 
gauge group and one
matter hypermultiplet can be generated by the following brane 
configuration: three parallel 5-branes
(labelled $\a=0,1,2$), the middle one  ($\a=1$) on top of an 
orientifold 6-plane, with $N$ 4-branes
stretched across from $\a=0$ through $\a=2$. The 6-orientifold at 
$\a=1$ enforces the condition that
the configuration be left/right-symmetric with respect to the 5-brane
 at site $\a=1$. This implies
that the branching index $B$ is 3 at all branching points. When the 
orientifold 6-plane carries
RR charge +4, the matter hypermultiplet turns out to transform in 
the symmetric representation of
$SU(N)$, while it transforms in the antisymmetric if it carries 
charge $-4$. We will analyse these
two cases separately.

We first consider the symmetric representation. The surface is given
 by \cite{KL}
\be
F(t,v)  = v^2\,t^3 +   f(v)\, t^2 + (-1)^N\,\Lambda^{N-2}\,g(v)\,t 
+ \Lambda^{3N-6}\, v^2=0,
\label{eq:40}
\ee
where 
\bea
f(v) & = & \prod_{i=1}^N (v-a_i)=\sum_{j=0}^N (-1)^j\,u_j\, v^{N-j}
\nonumber \\ 
g(v) & = & \prod_{i=1}^N (v+a_i)=\sum_{j=0}^N u_j\, v^{N-j}.
\label{eq:41}
\eea
The $u_j$ for $j=2,3,\ldots, N$ are a set of moduli parametrising 
the Coulomb branch, while $u_1$
is proportional to the bare mass of the hypermultiplet, and 
$u_0=1$.

The surface $\Sigma_g$ defined by equations (\ref{eq:40}) and 
(\ref{eq:41}) has genus $g=3N-2$, as
one finds by application of the Riemann--Hurwitz formula of equation 
(\ref{eq:12a}). On the
other hand, the dimension of the Coulomb branch is $N-1$. The full 
Jacobian of
$\Sigma_g$ contains a Prym subvariety that is invariant under the 
involution $\sigma\colon
(v,t)\rightarrow(-v,\Lambda^{2N-4}/t)$ of the surface 
\cite{KL}. One can check that the SW
differential given in equation (\ref{eq:14}) is invariant under 
$\sigma$, \ie,
$\sigma(\lambda_{SW})=\lambda_{SW}$. Hence the holomorphic 
differentials obtained by modular
differentiation of $\lambda_{SW}$ will also be invariant under
$\sigma$. A basis of such $\sigma$-invariant  differentials can  be
obtained with the help of  equations (\ref{eq:14}), (\ref{eq:40}) 
and (\ref{eq:41}). One finds
\be 
\omega_j=:{\p \lambda_{SW}\o\p u_j}= -{1\o F_t}\, \Big[(-1)^j\, t + 
(-1)^N\, \Lambda^{N-2}\Big] 
\,v^{N-j}\, \d v + \d (*), \quad j=2,3,\ldots, N.
\label{eq:42}
\ee 

As in previous sections, let us try to define an algebra of 
holomorphic differentials following equation (\ref{eq:8}). We first
observe from the definition of $F(t,v)=0$ that there are two
apparently inequivalent choices for the term to be modded out, 
namely
$f'(v)/(F_t)^2$ and
$g'(v)/(F_t)^2$. In fact these two choices are related by a moduli 
redefinition, $u_j\rightarrow (-1)^j\, u_j$, so they are not
independent. We tentatively set the product $\omega_i\, \omega_j$
equal to 
\be
\omega_i\, \omega_j=C_{ij}^l\,\omega_l\,\omega\quad {\rm mod}\, 
{f'(v)\o (F_t)^2}\,(\d v)^2
\label{eq:43}
\ee 
and examine whether or not the above equation can uniquely define a 
set of structure constants
$C_{ij}^l$. The left-hand side of equation (\ref{eq:43}) carries a 
$t$-dependence given by
$[(-1)^it+(-1)^N\,\Lambda^{N-2}]$ $[(-1)^jt+(-1)^N\,
\Lambda^{N-2}]$. None of these terms can be cancelled against 
the
prefactor $[(-1)^lt+(-1)^N\,\Lambda^{N-2}]$ of $\omega_l$, as 
there
is a summation over $l$ on the right-hand side of equation
(\ref{eq:43}). As a consequence, if the algebra is to hold, 
then
equation (\ref{eq:43}) must be understood as a polynomial in the 
two
variables $t$ and $v$, once the common factors $(\d v)^2/(F_t)^2$ 
have been cleared. In particular, the left-hand side has degrees
${\rm deg}_t(\omega_i\, \omega_j)=2$ and ${\rm deg}_v(\omega_i\,
\omega_j)=2N-4$ in
$t$ and $v$,   respectively, while those of $f'(v)$ are ${\rm deg}
_t(f'(v))=0$  and ${\rm
deg}_v(f'(v))=N-1$. Let $q_{ij}(t,v)$ denote the polynomial 
multiplying term modded out on the
right-hand side. Then we have ${\rm deg}_t(q_{ij}(t,v))=2$ and 
${\rm deg}_v(q_{ij}(t,v))=N-3$. A
straightforward computation gives $4N-7$ as the total number of 
coefficients to be determined if the
algebra is to hold. On the other hand, the number of available 
equations obtained by identification
of two polynomials in $t$ and $v$ with respective degrees $2$ and 
$2N-4$ is $6N-9$. We have an
overdetermined system of equations. The algebra of holomorphic 
differentials does not exist as
defined in equation (\ref{eq:43}).

It is in fact no surprise that we have not been able to define an 
algebra of holomorphic
differentials following the hyperelliptic pattern of equation 
(\ref{eq:8}). Not only is the surface 
$\Sigma_g$ defined by equations (\ref{eq:40}) and (\ref{eq:41}) 
non-hyperelliptic; it also cannot be
understood as a superposition of hyperelliptics. This conclusion can 
be arrived at by a counting
of moduli, or by the following argument.

Given that the branching index  $B$ is always 3,  completing three 
loops around any one branching
point $v_0$ on the base ${\bf CP}^1$ takes us from sheet $\a=0$, 
through sheet $\a=1$, to sheet
$\a=2$. This property is reflected in the presence of the prefactor 
$[(-1)^j\, t + (-1)^N\,
\Lambda^{N-2}]$ in the differential $\omega_j$ of equation 
(\ref{eq:42}). This prefactor is no
longer a monomial in $t$, as was the case for the models of sections 
3.1 and 3.2. Rather, it is a sum
of two monomials in  $t$. The one of order $t^1$ can be understood 
as being associated with sheets
$\a=0$ and $\a=1$, while that of order $t^0$ can be assigned to 
sheets $\a=1$ and $\a=2$. 

Next let us apply $\p/\p t$ to this brane configuration. The 
resulting surface,
\be 
{\p F\o \p t}= 3v^2\, t^2 + 2\,f(v)\, t + (-1)^N\,\Lambda^{N-2}\, 
g(v)=0,
\label{eq:46}
\ee
corresponds to an $SU(N-1)$ gauge theory with $N$ {\it fundamental}\/
 flavours
\cite{FARAGGI}, as a counting of powers of $v$ and $t$ reveals. Its 
genus is $g=N-2$. We also
observe that the power $N-2$ to which the quantum scale $\Lambda$ is 
raised is indeed the correct
one for an $SU(N-1)$ gauge theory with $N$  matter hypermultiplets 
in the fundamental representation.
It is known \cite{FARAGGI} that a reduction in the rank of the gauge 
group can be achieved by taking
the double scaling limit. However, we are not taking this limit 
here. Differentiation with respect to
$t$ removes the 5-brane at site $\a=2$, while $N$ semi-infinite 
4-branes remain to its right. The
latter account for the $N$ fundamental hypermultiplets, but the $N$ 
remaining 4-branes between the
5-branes at sites $\a=0$ and $\a=1$ are in excess for a gauge group  
$SU(N-1)$. 

This inconsistency between the brane configuration corresponding to 
the surface $\p F/\p t=0$ and 
the SW model that it actually describes can be easily interpreted. It 
is a consequence of the fact
that the initial surface $F=0$ cannot be understood as a 
superposition of hyperelliptics. The
basic building block of the original model consisted of three 
5-branes, plus one 6-orientifold on
top of the middle 5-brane to enforce a left/right symmetry with 
respect to $\a=1$. The effect of
this symmetry on the surface $\Sigma_g$ is to enforce a constant 
branching index $B=3$, so this
property is lost when one 5-brane is removed.  However, we  should 
emphasise that our arguments do
not prevent the existence of a non-hyperelliptic  associative 
algebra, according to the  pattern of
equation (\ref{eq:4}).

As a final example we will consider the antisymmetric representation 
of $SU(N)$. This case is very
similar to the previous one, so we will briefly report the final 
results.  The surface is
described   by
\cite{KL}
\be
F(t,v)=t^3+  \Big[ v^2\,f(v)+ 3\Lambda^{N+2}\Big]\, t^2 + \Lambda^
{N+2}\,
\Big[(-1)^N\,v^2\,g(v)+3\Lambda^{N+2}\Big]\,t + \Lambda^{3N+6}=0,
\label{eq:44}
\ee
with $f(v)$ and $g(v)$ given in equation (\ref{eq:41}). As in the 
symmetric representation, there
is an involution $\sigma$ of the surface \cite{KL} that is 
automatically taken account of when
considering modular derivatives of the SW differential, since 
$\sigma(\lambda_{SW})=\lambda_{SW}$.
A basis of $\sigma$-invariant holomorphic differentials on the above 
surface  is found to be
\be
\omega_j=:{\p \lambda_{SW}\o\p u_j}= -{1\o F_t}\, \Big[(-1)^j\, t + 
(-1)^N\, \Lambda^{N+2}\Big] 
\,v^{N+2-j}\, \d v + \d (*), \quad j=2,3,\ldots, N.
\label{eq:45}
\ee 
In trying to define an algebra of holomorphic differentials as in 
equation (\ref{eq:43}), by
modding out a term in $f'(v)\, (\d v)^2/(F_t)^2$, one again finds
that the prefactors $[(-1)^j\, t + (-1)^N\,\Lambda^{N+2}]$ in the
forms $\omega_j$ cannot be cancelled. Therefore the algebra, if it
exists, must be defined as a polynomial equation in the two 
variables
$t$ and $v$. The number of undetermined coefficients turns out to 
be
$4N+5$, while that of available equations is $6N+3$. Again we have 
an
overdetermined system of equations. Conclusions analogous to those
that were found for the symmetric representation continue to hold 
for
the antisymmetric representation as well.

\subsection{Models with 6-branes}

In this section we extend our analysis to the models of \cite{W} 
that include 6-branes. The gauge
group is $\prod _{\a=1}^n SU(k_{\a})$,  with matter hypermultiplets 
transforming in the sum of
bifundamental  representations $\sum_{\a=1}^{n-1}({\bf k}_{\a},\bar 
{\bf k}_{\a+1})$. Starting from
the same brane configuration as in section 3.1, we place $d_{\a}$ 
6-branes between the 5-branes at
sites $\a-1$ and $\a$. This adds $d_{\a}$ hypermultiplets in the 
fundamental represetation of
$SU(k_{\a})$. According to \cite{W}, the family of surfaces 
$\Sigma_g$ describing the Coulomb branch
of this theory is
\be 
F(t,v)=\sum_{\a=0}^{n+1}p_{k_{\a}}(v)\, \prod_{s=1}^{\a-1}J_s^{\a-s}
(v)\,t^{n+1-\a}=0,
\label{eq:30}
\ee 
where $p_{k_{\a}}(v)$ is given in equation (\ref{eq:11}). The 
polynomials $J_s(v)$ vanish
(with multiplicity 1) at the projections $e_a$ on the base 
${\bf CP}^1$ of the $d_{\a}$ 6-branes
that are located between sites $\a-1$ and $\a$, \ie, 
\be 
J_{\a}(v)=\prod_{a=1}^{d_{\a}}(v-e_{a}).
\label{eq:31}
\ee 

Now, on the above surface, the branchings between sheets  are not 
only effected by the
4-branes, but also by the 6-branes \cite{W}. A 6-brane placed 
between sheets $\a-1$ and $\a$
effects a multiple branching of {\it all}\/ sheets with $\b\geq\a$; 
the number of the latter is
therefore equal to the corresponding branching index. The $v$ 
coordinate of such a branching point
on the base ${\bf CP}^1$ is given by the value $e_{a}$ of the 
corresponding 6-brane.

Let us compute the genus $g$ of the surface defined by equation 
(\ref{eq:30}). For the sake of
simplicity we will assume that, on the base ${\bf CP}^1$,  no 4-brane
 ever has the same $v$
coordinate as a 6-brane. This simplifying assumption allows us to 
write the Riemann--Hurwitz formula
of equation (\ref{eq:12a}) as
\be
g={1\o 2}\, \sum_{p\in\Sigma_g}(B^{(4)}(p)-1) + {1\o 2}\, 
\sum_{p\in\Sigma_g}(B^{(6)}(p)-1) - n,
\label{eq:32}
\ee
where the superindices $(4)$ and $(6)$ indicate that the branching 
is effected by a 4- or a
6-brane, respectively. We immediately see that the genus $g$ is 
greater than the one given in
equation (\ref{eq:12}). When some of the $d_{\a}$ are non-vanishing, 
there is a non-zero
contribution $g^{(6)}$, 
\be
g^{(6)}={1\o 2}\, \sum_{p\in\Sigma_g}(B^{(6)}(p)-1),
\label{eq:33}
\ee
to add to the amount $g^{(4)}$ contributed by the 4-branes,
\be
g^{(4)}={1\o 2}\, \sum_{p\in\Sigma_g}(B^{(4)}(p)-1) - n=
\sum_{\a=1}^n (k_{\a}-1),
\label{eq:34}
\ee
so that $g=g^{(4)}+ g^{(6)}$. In order to compute the contribution 
$g^{(6)}$ we will make the
assumption, to be justified presently, that {\it no}\/ two 6-branes 
ever have the same $v$
coordinate on the base ${\bf CP}^1$. Then it suffices to know how 
many 6-branes we have,  plus how
many 5-branes are placed to the right of any given 6-brane. The 
first number tells us how many
branchings are effected by 6-branes; the second one tells us the 
corresponding branching index. The
precise value of $g^{(6)}$ so obtained is however immaterial to the 
discussion. It suffices to know
that $g^{(6)}>0$ except when all the $d_{\a}$ vanish. The 
contribution
$g^{(6)}$ also vanishes in the limiting case that $d_{\a}=0$ for all
$\a=1,2,\ldots, n-1$ and
$d_n\neq 0$. Then all the 6-branes are located between the last two 
sheets of $\Sigma_g$, so they
don't effect any branchings at all. Physically, this is equivalent 
to having $d_{\a}=0$ for all
$\a=1,2,\ldots, n$ and having all the 6-branes between the last two 
sheets replaced with the same
number of semi-infinite 4-branes, but now placed to the right of the 
last sheet \cite{W}.

In the presence of more than one 6-brane with the same value of $v$,
 the contribution $g^{(6)}$ of
the 6-branes to the overall genus $g$ decreases. Let us for 
simplicity take two 6-branes, one
placed between sites $\a-1$ and $\a$ in the chain of 5-branes, 
the other one between sites $\b-1$ and
$\b$. Without loss of generality we can assume $\a \leq \b
\leq n$. Denote their respective
projections on the base ${\bf CP}^1$ by $e_{\a}$ and $e_{\b}$. When 
$e_{\a}\neq e_{\b}$, we have one
branching point at $v=e_{\a}$ with branching index $B_{\a}=n+1-\a$, 
and another one at $v=e_{\b}$
with branching index $B_{\b}=n+1-\b$. The summand $B(p)-1$ in 
$g^{(6)}$ therefore
receives a contribution $2n-(\a+\b)$. On the other hand, when 
$e_{\a}=e_{\b}$, the two branching
points have melted into one, with branching index $B_{\a}=n+1-\a$, 
and a contribution of $n-\a$
to the summand $B(p)-1$ of $g^{(6)}$ in equation (\ref{eq:33}). As 
$n-\a\leq 2n-(\a+\b)$, we see
that the contribution $g^{(6)}$ decreases. The equality holds if and 
only if $\b=n$; this
correponds to the trivial case mentioned above, \ie, when one of the 
6-branes is
placed between the last two sheets of $\Sigma_g$.

The mechanism just described, whereby two 6-branes are made to have
 coincident $v$-projections on
the base ${\bf CP}^1$,  corresponds to a transition to a Higgs phase 
\cite{W}. We observe that the
models of section 3 did not exhibit this behaviour. As explained 
there, any one 4-brane between
sites $\a-1$ and $\a$ could be made to have the same $v$-projection 
as any other 4-brane between
sites $\a$ and $\a+1$, with the genus $g$ remaining constant. In 
section 3, the decrease in the
number of branching points when two or more 4-branes had the same 
$v$-projection was compensated by
an increase in the branching index. This was due to the structure 
of the surface $\Sigma_g$ as a
superposition of hyperelliptics: no single 4-brane ever connected
more than two adjacent sheets. We observe that the surfaces of  this
section no longer enjoy this property, due to the branchings effected
by the 6-branes.

Hence, by differentiation of the SW differential $\lambda_{SW}$ with
respect to the moduli $u_j^{(\a)}$ we do not obtain a complete basis
of holomorphic  differentials on this surface (unless all the
$d_{\a}=0$). All that one obtains is a basis for the $g^{(4)}$
holomorphic differentials that can be associated with the 4-branes.
The property that the surface $\Sigma_g$ ceases to be a superposition
of hyperelliptics is reflected in the appearance of new moduli in the
theory, other than those associated with the order  parameters
$u_j^{(\a)}$ on the Coulomb branch. 

We observe from \cite{W} that the 4-manifold ${\bf Q}$ in which the 
surface $\Sigma_g$ is immersed  is no longer the space of section 3.
On ${\bf R}^3 \times {\bf S}^1$, the hyperelliptic involution
$\sigma$ was a discrete transformation that squared to unity. In the
presence of 6-branes, the 4-manifold ${\bf Q}$ becomes
multi--Taub--NUT space \cite{Q}.  On the latter  there is a
continuous ${\bf C}^*$-action given by the complexification of a
$U(1)$ rotation symmetry around the ${\bf S}^1$ direction of
$M$-theory \cite{W}. The hyperelliptic involution on
${\bf R}^3\times {\bf S}^1$ is in fact a discrete remnant (in the
limiting case when all the $d_{\a}$ vanish) of this ${\bf
C}^*$-action on multi--Taub--NUT space.

Therefore, if $d_{\a}\neq 0$ for some $\a<n$, we are 
missing 
$g^{(6)}$ holomorphic differentials, so we cannot define an algebra.
However, we can draw some  conclusions. Given that the structure of
$\Sigma_g$ isn't a superposition of  hyperelliptics, if any
algebra of holomorphic differentials is to hold at all, we do not
expect it to  conform to the hyperelliptic pattern of equation
(\ref{eq:8}).  On physical grounds, the number
$d_{\a}$ of 6-branes placed between sites $\a-1$ and $\a$ is a free 
parameter that one can vary, in order to obtain a theory with a
vanishing beta function \cite{W}. A  vanishing beta function
indicates the existence of a new modulus in the theory, namely, the 
gauge coupling constant. From
\cite{RUSOS} we do not expect the WDVV equation to hold in this  
case.

\section{Summary and conclusions}

Using $M$-theory techniques, large classes of ``new" SW models have 
been constructed recently whose
moduli spaces (in their Coulomb branches) are described by 
non-hyperelliptic Riemann surfaces.
``Old" SW models (\ie, prior to the advent of $M$-theory and 
geometric engineering) were typically
described by hyperelliptic Riemann surfaces, with their corresponding
 prepotentials satisfying the
WDVV equation. In this paper we have posed the question of whether 
or not the prepotentials
associated with these new SW models continue to satisfy the WDVV 
equation, despite the loss of
the property of hyperellipticity of the corresponding Riemann 
surfaces. The answer to this question
comes in two steps. One first needs to define an associative algebra 
for the holomorphic
1-forms on the Riemann surface. Next one expresses the (third 
derivatives of the) 
prepotential in terms of those differentials (the so-called residue 
formula).  Associativity of
the algebra of 1-forms is then an equivalent statement of the 
validity of the WDVV equation.

We have taken the first of the two steps mentioned above, deferring a 
proof of the residue
formula for an upcoming publication. We find two substantially 
different classes of
non-hyperelliptic SW models. In the first one, it turns out to be 
possible to define an associative
algebra of holomorphic differentials. Although non-hyperelliptic, 
these Riemann surfaces can be
understood (roughly speaking) as a superposition of hyperelliptic 
building blocks. The
construction of these surfaces is carried out in such a way that all 
properties of the hyperelliptic
building block pertaining to the algebra of holomorphic 1-forms are 
maintained. Characteristically,
the SW models so described correspond to product gauge groups, with 
matter hypermultiplets
transforming in (sums of) bifundamental representations.  The loss of
 hyperellipticity in these
models is of no import, and therefore their prepotentials can be 
expected to satisfy the WDVV
equation, much as their hyperelliptic ancestors did. From the 
viewpoint of their $M$-theory
construction, these theories involve 4- and 5-branes only (plus, 
possibly, 4-orientifolds as well).
We observe that, in these cases, $\Sigma_g$ is a surface in the 
4-manifold ${\bf Q}= {\bf
R}^3\times {\bf S}^1$. All the moduli in the theory are those 
associated with the physical order
parameters on the Coulomb branch. The latter are determined solely 
by the gauge group; no new
moduli appear in the passage from the ``old" hyperelliptic  SW 
models to these
``new" non-hyperelliptic cases.

This allows us to formulate a sufficient condition for the algebra 
of holomorphic differentials to
be associative. Namely, if the surface $\Sigma_g$ can be decomposed 
as a superposition of
hyperelliptics (in the manner described in the body of the paper)  
then an associative algebra of
holomorphic differentials will hold. Whether or not this condition 
is also necessary remains an
open question.

A second class of non-hyperelliptic SW models is analysed, in which 
it turns out to be impossible
to define an associative algebra of holomorphic differentials 
following the pattern of the
hyperelliptic case. We would like to underline the fact that this 
does not rule out the possibility
of defining an associative algebra. However, such an algebra (if it 
exists at all) will have to
conform to the non-hyperelliptic pattern established in section 
2.1. Geometrically, it  is observed
already at the level of the corresponding Riemann surfaces that 
hyperellipticity is lost in a more
fundamental way, because it is no longer  possible to ``decompose" 
the surface as a superposition of
hyperelliptic building blocks. From a physical viewpoint, their 
$M$-theory construction requires the
inclusion of 6-branes and/or 6-orientifolds. 

Specifically, in the presence of 6-orientifolds, the branching index 
is compelled to take
fixed values greater than 2. The increase in the value of the genus 
(with respect to the
hyperelliptic case with the same number of moduli) is compensated by 
a restriction to a
certain Prym subvariety.  The dimension of the latter is twice the 
number of independent
moduli. However, even after this restriction, an associative algebra 
following the pattern of the
hyperelliptic case is not possible. Typically, these SW models 
describe gauge theories with matter
hypermultiplets in representations higher than the fundamental.

In the presence of 6-branes, the loss of hyperellipticity is more 
profound, because it can be
ascribed to the appearance of new moduli. For example, the number of
 6-branes included in
the brane configuration can be fine-tuned in such a way that the 
beta function vanishes. The gauge
coupling constant then becomes a modulus. Even before reaching that 
critical value in the
number of 6-branes, when the beta function continues to be negative,
 the 6-branes cause an
increase in the value of the genus with respect to the case with the
 lowest value of $g$ that is
compatible with the same number of moduli, \ie, the case with no 
6-branes at all. This
increase cannot be compensated by restricting to a certain Prym 
subvariety. In consequence, modular
derivatives of the SW differential no longer provide us with a 
complete basis of holomorphic
differentials. This can be rephrased by saying that we are missing 
moduli, so we cannot write
down a complete basis of differentials. In these cases we observe 
that $\Sigma_g$ is a surface in the
4-manifold ${\bf Q}$ given by a multi--Taub--NUT  space. A natural 
question to ask is whether or
not the 4-manifold ${\bf Q}$ can provide the missing moduli.

Our analysis reveals a connection between the algebra of holomorphic 
differential forms on the
Riemann surface and the configuration of $M$-theory branes used in 
the construction of the
corresponding SW model. We hope these observations may provide some 
insight into a purely
$M$-theoretic derivation of the WDVV equation, \ie, one without 
recourse to an underlying algebra of
differentials. Looking beyond, one could pose the question of whether
 or not there is some
generalisation of the WDVV equation that would hold in the models 
examined in section 5.

We hope the observations made here may prove useful in clarifying 
these issues.

\noindent{\bf Acknowledgements}

It is a great pleasure to thank M. Matone and M. Tonin for 
discussions and encouragement. This work
has been supported by the Commission of the European Community 
under contract FMRX-CT96-0045.

\begin{center} 
{\bf Appendix }
\end{center}

Below we present a proof of equations (\ref{eq:17}), (\ref{eq:19})
and (\ref{eq:20}). 

Let us start with equation (\ref{eq:17}). We have observed that the
1-form $\phi_j^{(n)}$ is holomorphic on the surface $F=0$, because 
it
coincides with $\p \lambda_{SW}/\p u_j^{(n)}$.  We also know that 
its
zeroes will be at the points on the surface $F=0$ lying above $v=0$
and
$v=\infty$ on the base ${\bf CP}^1$.  In fact, from the observation 
made after equation
(\ref{eq:21a}), the zeroes of $\phi_j^{(n)}$ will lie on the sheets 
$\a=n$ and $\a=n+1$ of the
surface $F=0$. We have also made the assumption that neither $v=0$ 
nor $v=\infty$ are branching
points of $F=0$. From here we conclude a divisor $[\phi_j^{(n)}]$ of 
the general form
$$
[\phi_j^{(n)}]= c_1\, (0_{n-1} + 0_n) + c_2\, (\infty_{n-1} + 
\infty_n),
$$
where $c_1$ and $c_2$ are certain integers.  Again, the fact that 
$v=0$ is not a branching point of
$F=0$, together with equation (\ref{eq:16}), dictates that $c_1=
k_n-j$. It now suffices  to use
equation (\ref{eq:13}) in order to conclude that the remaining 
coefficient $c_2$ is
$\sum_{l=1}^{n-1}k_l + j -(n+1)$, as stated in equation 
(\ref{eq:17}).

Equation (\ref{eq:19}) is proved along the same lines as 
(\ref{eq:17}). We start from the
observation that, for every fixed value of $\a=n-1, n-2, \ldots, 1$, 
the 1-form $\phi_j^{(\a)}$ is
holomorphic on the surface $\p^{n-\a} F/\p t^{n-\a}=0$. From here 
conclude that the divisor
$[\phi_j^{(\a)}]$ on  $\p^{n-\a} F/\p t^{n-\a}=0$ is as stated in 
equation (\ref{eq:19}).
The applicable value of the genus is now $g_{\a}=\sum_{l=1}^{\a}
(k_l-1)$.

{}Finally, in proving equation (\ref{eq:20}) we start from the 
observation that $v=0$ has been
assumed not to be a branching point of $\p^{n-\a} F/\p t^{n-\a}=0$ 
for any $\a=1,2,\ldots, n$ (the
value $\a=n$ corresponding by convention to the surface $F=0$). 
Hence the coefficient $c_1=k_{\a}-j$
multiplying $(0_{\a-1}+0_{\a})$ in the divisor $[\phi_j^{(\a)}]$ is 
correct not only on $\p^{n-\a}
F/\p t^{n-\a}=0$ for $\a=n-1, n-2, \ldots, 1$, as per equation 
(\ref{eq:19}), but also on $F=0$. 
No new poles or zeroes appear when extending the 1-form 
$\phi_j^{(\a)}$ from $\p^{n-\a} F/\p
t^{n-\a}=0$ to $F=0$, except possibly at $v=\infty$, \ie, at 
$\infty_{\a-1}$ and $\infty_{\a}$, for
the same reasons as previously. Hence all that remains to determine 
is the coefficient $c_2$ in
front of the term  $(\infty_{\a-1} + \infty_\a)$. This is again 
fixed by the requirement in equation
(\ref{eq:13}), after observing that the applicable value of the 
genus is now $g=g_n=\sum_{l=1}^n
(k_l-1)$. Hence the divisor in equation (\ref{eq:20}) is correct. 
Now, holomorphicity of
$\phi_j^{(\a)}$ on $F=0$ is equivalent to the requirement that both 
coefficients in (\ref{eq:20}) be
positive. This happens if, and only if, for every $\a$ in the range 
$1,2,\ldots, n$, we have that
$j$ runs over the range $2,3,\ldots, k_{\a}$. Indeed, from the 
coefficient $c_1$ of
$(0_{\a-1}+0_{\a})$ we obtain the condition that $k_{\a}\geq j$. 
Next let $j\geq 2$. Then, from the
coefficient $c_2$ of $(\infty_{\a-1} +\infty_\a)$, we have that $j+
 \sum_{l\neq\a} ^n k_l \geq
2+\sum_{l\neq\a} ^n 1=2+(n-1)=n+1$, so also this $c_2$ is positive. 
This finally establishes the
holomorphicity of $\phi_j^{(\a)}$ on $F=0$.

\end{document}